\documentclass[journal,superscriptaddress,11pt,draftclsnofoot,onecolumn,showpacs]{revtex4-1}
\usepackage{amsmath}

\usepackage{graphicx}
\usepackage{dcolumn}
\usepackage{bm}
\usepackage{esdiff}
\usepackage{natbib}
\usepackage[colorlinks=true,citecolor=blue,linkcolor=blue,pdfhighlight =/O]{hyperref}
\usepackage{lineno}
\hypersetup{urlcolor=blue}

\begin{document}
\title{ \Large Anomalies of upper critical field in the spinel superconductor LiTi$_2$O$_{4-\delta}$}

\author{Zhongxu Wei}
\altaffiliation{These authors contributed equally to this work.}
\affiliation{Beijing National Laboratory for Condensed Matter Physics, Institute of Physics, Chinese Academy of Sciences, Beijing 100190, China.}
\affiliation{School of Physical Sciences, University of Chinese Academy of Sciences, Beijing 100049, China.}

\author{Ge He}
\altaffiliation{These authors contributed equally to this work.}
\affiliation{Beijing National Laboratory for Condensed Matter Physics, Institute of Physics, Chinese Academy of Sciences, Beijing 100190, China.}
\affiliation{School of Physical Sciences, University of Chinese Academy of Sciences, Beijing 100049, China.}

\author{Wei Hu}
\altaffiliation{These authors contributed equally to this work.}
\affiliation{Beijing National Laboratory for Condensed Matter Physics, Institute of Physics, Chinese Academy of Sciences, Beijing 100190, China.}
\affiliation{School of Physical Sciences, University of Chinese Academy of Sciences, Beijing 100049, China.}

\author{Zhongpei Feng}
\affiliation{Beijing National Laboratory for Condensed Matter Physics, Institute of Physics, Chinese Academy of Sciences, Beijing 100190, China.}
\affiliation{School of Physical Sciences, University of Chinese Academy of Sciences, Beijing 100049, China.}

\author{Jie Yuan}
\affiliation{Beijing National Laboratory for Condensed Matter Physics, Institute of Physics, Chinese Academy of Sciences, Beijing 100190, China.}
\affiliation{Songshan Lake Materials Laboratory, Dongguan, Guangdong 523808, China}

\author{Chuanying Xi}
\affiliation{Anhui Province Key Laboratory of Condensed Matter Physics at Extreme Conditions, High Magnetic Field Laboratory of the  Chinese Academy of Science, Hefei 230031, Anhui, China.}

\author{Qian Li}
\affiliation{Beijing National Laboratory for Condensed Matter Physics, Institute of Physics, Chinese Academy of Sciences, Beijing 100190, China.}
\affiliation{Songshan Lake Materials Laboratory, Dongguan, Guangdong 523808, China}

\author{Beiyi Zhu}
\affiliation{Beijing National Laboratory for Condensed Matter Physics, Institute of Physics, Chinese Academy of Sciences, Beijing 100190, China.}
\affiliation{Songshan Lake Materials Laboratory, Dongguan, Guangdong 523808, China}

\author{Fang Zhou}
\affiliation{Beijing National Laboratory for Condensed Matter Physics, Institute of Physics, Chinese Academy of Sciences, Beijing 100190, China.}
\affiliation{School of Physical Sciences, University of Chinese Academy of Sciences, Beijing 100049, China.}
\affiliation{Songshan Lake Materials Laboratory, Dongguan, Guangdong 523808, China}

\author{Xiaoli Dong}
\affiliation{Beijing National Laboratory for Condensed Matter Physics, Institute of Physics, Chinese Academy of Sciences, Beijing 100190, China.}
\affiliation{School of Physical Sciences, University of Chinese Academy of Sciences, Beijing 100049, China.}
\affiliation{Songshan Lake Materials Laboratory, Dongguan, Guangdong 523808, China}

\author{Li Pi}
\affiliation{Anhui Province Key Laboratory of Condensed Matter Physics at Extreme Conditions, High Magnetic Field Laboratory of the  Chinese Academy of Science, Hefei 230031, Anhui, China.}

\author{F. V. Kusmartsev}
\affiliation{Department of Physics, Loughborough University, Loughborough LE11 3TU, UK.}
\affiliation{Micro/Nano Fabrication Laboratory Microsystem and THz Research Center, Chengdu, Sichuan, China}
\affiliation{ITMO University, St. Petersburg 197101, Russia}

\author{Zhongxian Zhao}
\affiliation{Beijing National Laboratory for Condensed Matter Physics, Institute of Physics, Chinese Academy of Sciences, Beijing 100190, China.}
\affiliation{School of Physical Sciences, University of Chinese Academy of Sciences, Beijing 100049, China.}
\affiliation{Songshan Lake Materials Laboratory, Dongguan, Guangdong 523808, China}

\author{Kui Jin}\email[Corresponding author: ]{kuijin@iphy.ac.cn}
\affiliation{Beijing National Laboratory for Condensed Matter Physics, Institute of Physics, Chinese Academy of Sciences, Beijing 100190, China.}
\affiliation{School of Physical Sciences, University of Chinese Academy of Sciences, Beijing 100049, China.}
\affiliation{Songshan Lake Materials Laboratory, Dongguan, Guangdong 523808, China}

\date{\today}
\begin{abstract}
High-field electrical transport and point-contact tunneling spectroscopy were used to investigate superconducting properties of the unique spinel oxide, LiTi$_2$O$_{4-\delta}$ films with various oxygen content. We find that the upper critical field $B_\mathrm{c2}$ gradually increases as more oxygen impurities are brought into the samples by carefully tuning the deposition atmosphere. It is striking that although the superconducting transition temperature and energy gap are almost unchanged, an astonishing isotropic $B_\mathrm{c2}$ up to $\sim$ 26 Tesla is observed in oxygen-rich sample, which is doubled compared to the anoxic sample and breaks the Pauli limit. Such anomalies of $B_\mathrm{c2}$ were rarely reported in other three dimensional superconductors. Combined with all the anomalies, three dimensional spin-orbit interaction induced by tiny oxygen impurities is naturally proposed to account for the remarkable enhancement of $B_\mathrm{c2}$ in oxygen-rich LiTi$_2$O$_{4-\delta}$ films. Such mechanism could be general and therefore provides ideas for optimizing practical superconductors with higher $B_\mathrm{c2}$.
\end{abstract}

\maketitle

In conventional Bardeen-Cooper-Schrieffer (BCS) superconductors, time reversal and spatial inversion symmetries are essential to the formation of Cooper pairs. An external field $B$ violates the time reversal symmetry in a superconductor, thereby may break the Cooper pairs\cite{James1969}. Superconductivity will be completely destroyed when $B$ reaches an upper critical field $B_\mathrm{c2}$ for type-II superconductors. The underlying mechanisms that dominate and effectively increase $B_\mathrm{c2}$ intrigue intensive interests in condensed matter physics. Generally, $B_\mathrm{c2}$ is determined by orbital pair-breaking for conventional superconductors since spin susceptibility goes to zero at low temperatures\cite{Sohn2018}. In this case, $B_\mathrm{c2}$ can be enhanced by several factors that suppress orbital pair-breaking, such as narrow bands\cite{Holczer1991}, short mean free path\cite{Fischer1978}, and strong electron-phonon coupling\cite{Orlando1979}. Alternatively, spin flip induced by $B$ will dominate $B_\mathrm{c2}$ when orbital effect is eliminated such as ultrathin Be\cite{Adams1998} and Al films\cite{Tedrow1982} whose dimensionality is reduced. No matter which of aforementioned effects happens, $B_\mathrm{c2}$ is not expected to break the Clogston-Chandrasekhar limit\cite{Clogston1962,Chandrasekhar1962} or the Pauli paramagnetic limit $B_\mathrm{P}$, where $B_\mathrm{P} = 1.84 T_{\mathrm{c}}$ (in units of Tesla) for conventional superconductors and $T_\mathrm{c}$ is the superconducting transition temperature.

The upper critical field $B_\mathrm{c2}$ can be further enhanced and even exceed $B_\mathrm{P}$ with some unconventional mechanisms where spin paramagnetism is crucial in the superconducting state, such as Fulde-Ferrell-Larkin-Ovchinnikov states\cite{Matsuda2007}, spin triplet pairing\cite{Huy2007,Aoki2012} and spin-orbit interaction\cite{Herranz2015,Lu2015,Sohn2018}. In particular, spin-orbit interaction can be divided into two effects, i.e. spin-orbit coupling (SOC) and spin-orbit scattering (SOS). In the former case, the effect of spin flip can be weakened by locking spin parallel or perpendicular to in-plane due to Rashba SOC\cite{Herranz2015} or Zeeman SOC\cite{Lu2015,Sohn2018}, respectively. As for SOS, spin-flip scattering will mix the different spin states and lead to a finite Pauli paramagnetism even at zero temperature thus enhancing $B_\mathrm{c2}$\cite{Maki_I}. Generally, the effects of spin-orbit interaction are prominent in low dimensional materials with non-centrosymmetric lattice structure. As a result, it is rare to see the breaking of Pauli limit in conventional superconductors with centrosymmetry.

LiTi$_2$O$_{4-\delta}$ is a centrosymmetric conventional superconductor with highly nontrivial spin-orbit structure\cite{Jin2015}. For ideal LiTi$_2$O$_{4}$ structure, the valence of Ti is 3.5+ with low $d$-orbital filling and Jahn-Teller distortions, implying that large orbital freedom exists in this system\cite{Satpathy1987}. Orbital order and spin-orbit scattering are unveiled in the anoxic films where amounts of ordering oxygen vacancies have been explicated\cite{Jin2015,He2017_LTO}. As the vacancies are gradually filled by introducing oxygen atoms, the orbital order is suppressed locally and spin-orbit scattering is enhanced at low temperatures\cite{Jia2018}. Therefore, studying the evolution of $B_\mathrm{c2}$ of LiTi$_2$O$_{4-\delta}$ with various oxygen content may disclose some mechanisms of the spin-orbit interaction in centrosymmetric superconductors.

In this letter, we present systematic transport measurements and point-contact tunneling spectroscopy study of LiTi$_2$O$_{4-\delta}$ films with various oxygen content. For oxygen-rich samples, a significantly enhanced $B_\mathrm{c2}$ up to $26$~T at low temperatures is observed. Such doubled $B_\mathrm{c2}$, compared with anoxic samples, is isotropic. Intriguingly, superconducting energy gap and $T_\mathrm{c}$ are almost the same for all samples. Temperature-dependent $B_\mathrm{c2}$ can be well fitted by the Werthamer-Helfand-Hohenberg (WHH) theory considering spin paramagnetism and spin-orbit scattering\cite{Werthamer1966}, yet it leads to an underestimation of relaxation time for oxygen-rich samples. Here we consider the three dimensional spin-orbit interaction induced by oxygen impurities, which is beyond the WHH theory, thereby giving an universal understanding of the enhancement of $B_\mathrm{c2}$ and other anomalies.

High-quality LiTi$_2$O$_{4-\delta}$ thin films were epitaxially grown on MgAl$_2$O$_4$ (001) substrates by pulsed laser deposition under various oxygen pressures from basic vacuum ($< 1 \times 10^{-6}$~Torr) to $\sim 5 \times 10^{-6}$~Torr\cite{Jia2018}. The temperature- and field-dependence of resistance were measured by a standard four-probe method in PPMS-$16$T, and steady high magnetic field facility with field up to $33$~T. Point-contact measurements were performed by employing a homemade probe using Pt/Ir tips. The differential conductance $\mathrm{d}I/\mathrm{d}V$ is measured by standard lock-in technique in quasi-four-probe configuration.

\begin{figure}[ht!]
	\centering
	\includegraphics[width=\linewidth]{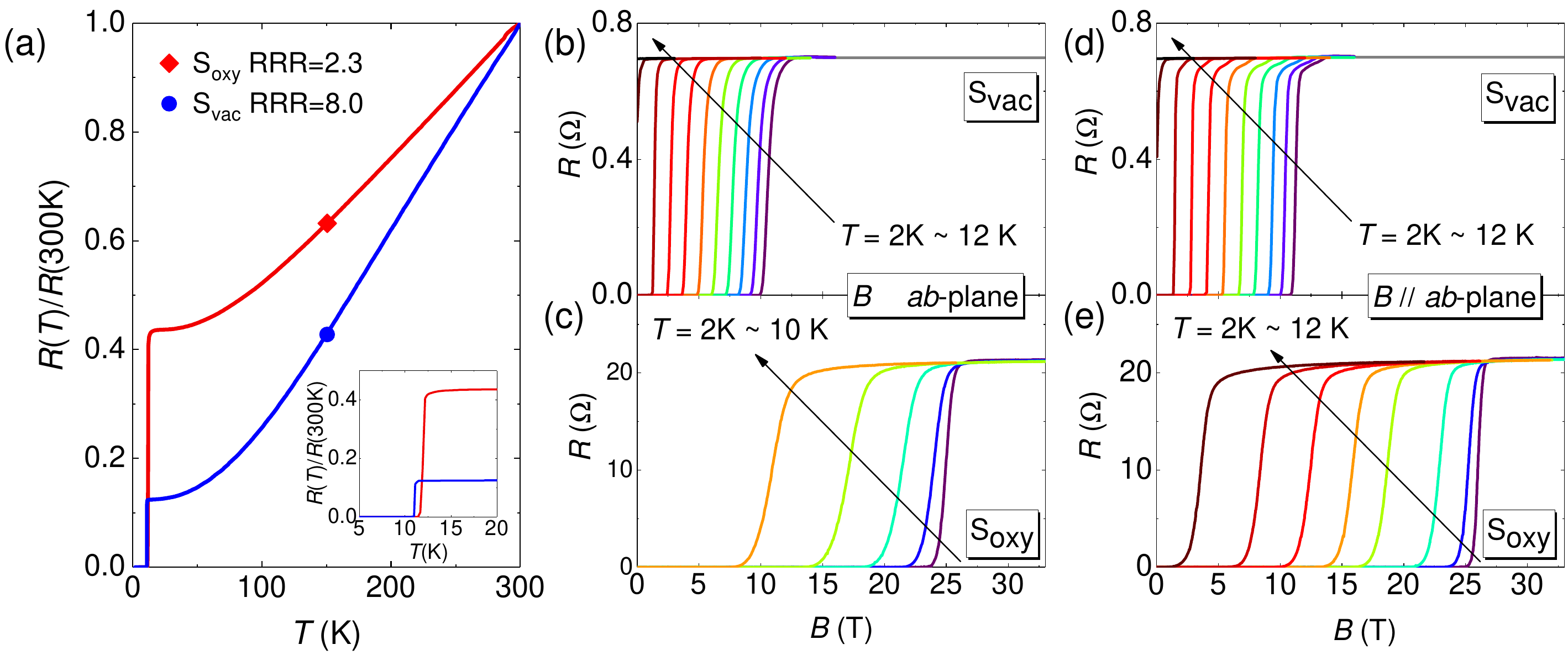}
	\caption{ \textbf{Temperature and magnetic field-dependent resistance of LiTi$_2$O$_{4-\delta}$ films.} (a)~$R(T)$ curves of S$_\mathrm{oxy}$ (red line) and S$_\mathrm{vac}$ (blue line). They have similar $T_\mathrm{c}$~($\sim$11~K) but different RRR, i.e. RRR = 2.3 for S$_\mathrm{oxy}$ and RRR = 8.0 for S$_\mathrm{vac}$. Inset: Zoom in of $R(T)$ curves. (b)~$R(B)$ curves of S$_\mathrm{vac}$ from $2$~K to $12$~K with $\Delta T = 1~K$. (c)~$R(B)$ curves of S$_\mathrm{oxy}$ from $2$~K to $10$~K with $\Delta T = 2~K$. The magnetic field is perpendicular to $ab$-plane of S$_\mathrm{vac}$ (b) and S$_\mathrm{oxy}$ (c). (d)~$R(B)$ curves of S$_\mathrm{vac}$ from $2$~K to $12$~K with $\Delta T = 1~K$. (e)~$R(B)$ curves of S$_\mathrm{oxy}$ at various temperatures. $T \in [2~K,8~K]$ with $\Delta T = 2~K$; $T \in [9~K,12~K]$ with $\Delta T = 1~K$. The magnetic field is parallel to the $ab$-plane of S$_\mathrm{vac}$ (d) and S$_\mathrm{oxy}$ (e). The grey lines in (b) and (d) are linearly extrapolated from experimental data.}
	\label{fig1}
\end{figure}

Fig.~\ref{fig1}~(a) shows the resistance versus temperature curves of S$_\mathrm{oxy}$ and S$_\mathrm{vac}$, deposited under oxygen pressure $\sim 5 \times 10^{-6}$~Torr and basic vacuum, respectively. Both samples display similar $T_\mathrm{c}$ of $11.5\pm0.5$~K with narrow transition widths. However, the residual resistance ratio (RRR), defined by room temperature resistance over resistance at $20$~K, shows remarkable difference, i.e. $8.0$ for S$_\mathrm{vac}$ and $2.3$ for S$_\mathrm{oxy}$. These results are consistent with our previous report\cite{Jia2018}. The magnetic field-dependent resistance $R(B)$ isotherms of S$_\mathrm{vac}$ and S$_\mathrm{oxy}$ with field perpendicular to the $ab$-plane of films are shown in Fig.~\ref{fig1}~(b) and~(c), respectively. It is intriguing that $B_\mathrm{c2}$ of S$_\mathrm{oxy}$ is prominently enhanced compared with that of S$_\mathrm{vac}$. In addition, the enhancement of $B_\mathrm{c2}$ is also observed when the magnetic field is parallel to the $ab$-plane, as shown in Fig.~\ref{fig1}~(d) and~(e).

\begin{figure}[ht!]
	\centering
	\includegraphics[width=\linewidth]{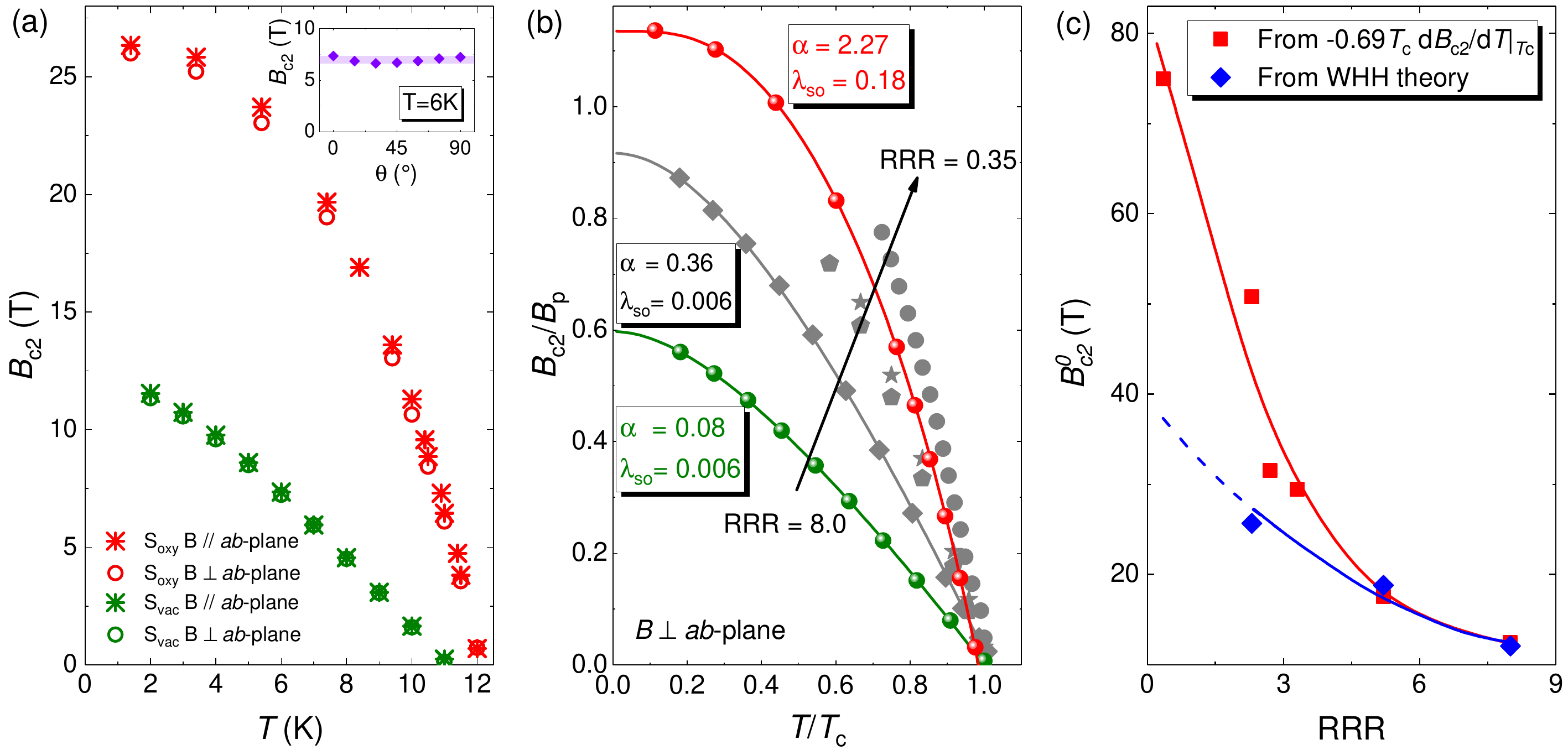}
	\caption{ \textbf{Upper critical field-temperature phase diagram of LiTi$_2$O$_{4-\delta}$ films.} (a)~The temperature-dependent $B_\mathrm{c2}$ of S$_\mathrm{oxy}$ and S$_\mathrm{vac}$ with field parallel and perpendicular to $ab$-plane respectively. Inset: Angle-dependent $B_\mathrm{c2}$ of S$_\mathrm{vac}$ at $6$~K. Here $\mathrm{\theta}$ stands for the angle between the applied filed and the $ab$-plane. (b)~The temperature-dependent normalized upper critical field $B_\mathrm{c2}/B_\mathrm{P}$ of series of LiTi$_2$O$_{4-\delta}$ films with RRR equal to 8.0, 5.2, 3.3, 2.7, 2.3 and 0.35. Some of data are fitted by the WHH theory (solid lines). $\lambda_\mathrm{so}$ and $\alpha$ are fitting parameters of the WHH theory. (c)~The values of $B^{0}_\mathrm{c2}$ of these samples, which are estimated by $-0.69 T_\mathrm{c} (\mathrm{d}B_\mathrm{c2} / \mathrm{d}T)|_{T_\mathrm{c}}$ (red squares) and the WHH theory (blue diamonds) respectively. The solid and dash lines are to guide the eye.}
	\label{fig2}
\end{figure}

In order to quantitatively study the enhancement of $B_\mathrm{c2}$, we extract the temperature-dependent $B_\mathrm{c2}$ of S$_\mathrm{vac}$ and S$_\mathrm{oxy}$ with field parallel and perpendicular to $ab$-plane, as shown in Fig.~\ref{fig2}~(a). The value of $B_\mathrm{c2}$ is evaluated at $90\%$ of the resistance transition relative to the normal state resistance. It is found that the $B_\mathrm{c2}$ of S$_\mathrm{oxy}$ at $2$~K is $\sim 26.0$~T, more than doubled compared to that of S$_\mathrm{vac}$ ($\sim 11.3$~T). Furthermore, the $B_\mathrm{c2}$ of LiTi$_2$O$_{4-\delta}$ is isotropic, which is confirmed by angle-dependent magnetoresistance measurements of S$_\mathrm{vac}$, as shown in the inset of Fig.~\ref{fig2}~(a). To get further insight into the isotropic enhancement of $B_\mathrm{c2}$, we choose RRR as a good quantity to distinguish different samples since RRR monotonically decreases with increasing the oxygen pressure\cite{Jia2018}. Fig .~\ref{fig2}~(b) shows normalized temperature ($T/T_\mathrm{c}$) dependence of $B_\mathrm{c2}/B_\mathrm{P}$ for samples with different RRR. Obviously, $B_\mathrm{c2}/B_\mathrm{P}$ gradually increases as RRR decreases, and the $B_\mathrm{c2}$ of S$_\mathrm{oxy}$ exceeds $B_\mathrm{P}$ at low temperatures. Typically, $B_\mathrm{c2}/B_\mathrm{P}$ versus $T/T_\mathrm{c}$ of three selected samples with much different RRR (i.e. $8.0$, $5.2$, and $2.3$) are fitted by the WHH theory\cite{Werthamer1966}. The theoretical fits match well with experimental data, considering Maki parameter $\alpha$ and spin-orbit scattering parameter $\lambda_\mathrm{so}$\cite{Maki_III,Werthamer1966}. $\alpha$ and $\lambda_\mathrm{so}$ are approximately zero in the case of large RRR and increase with decreasing RRR, tempting us to consider spin paramagnetism and spin-orbit scattering for the enhancement of $B_\mathrm{c2}$. The zero-temperature upper critical field $B_\mathrm{c2}^{0}$ extrapolated from the WHH fitting is shown in Fig.~\ref{fig2}~(c). In addition, a widely used formula which considers orbital effect only, $B_\mathrm{c2}^{0} = -0.69 T_\mathrm{c} (\mathrm{d}B_\mathrm{c2} / \mathrm{d}T)|_{T_\mathrm{c}}$, is also employed to estimate $B_\mathrm{c2}^{0}$, as shown in Fig.~\ref{fig2}~(c). The difference in $B_\mathrm{c2}^{0}$ by these two methods is small for samples with large RRR, while it becomes more and more prominent as RRR decreases, suggesting again that effects other than the orbital depairing should be considered for the $B_\mathrm{c2}$ anomalies.


\begin{figure}[ht!]
	\centering
	\includegraphics[width=\linewidth]{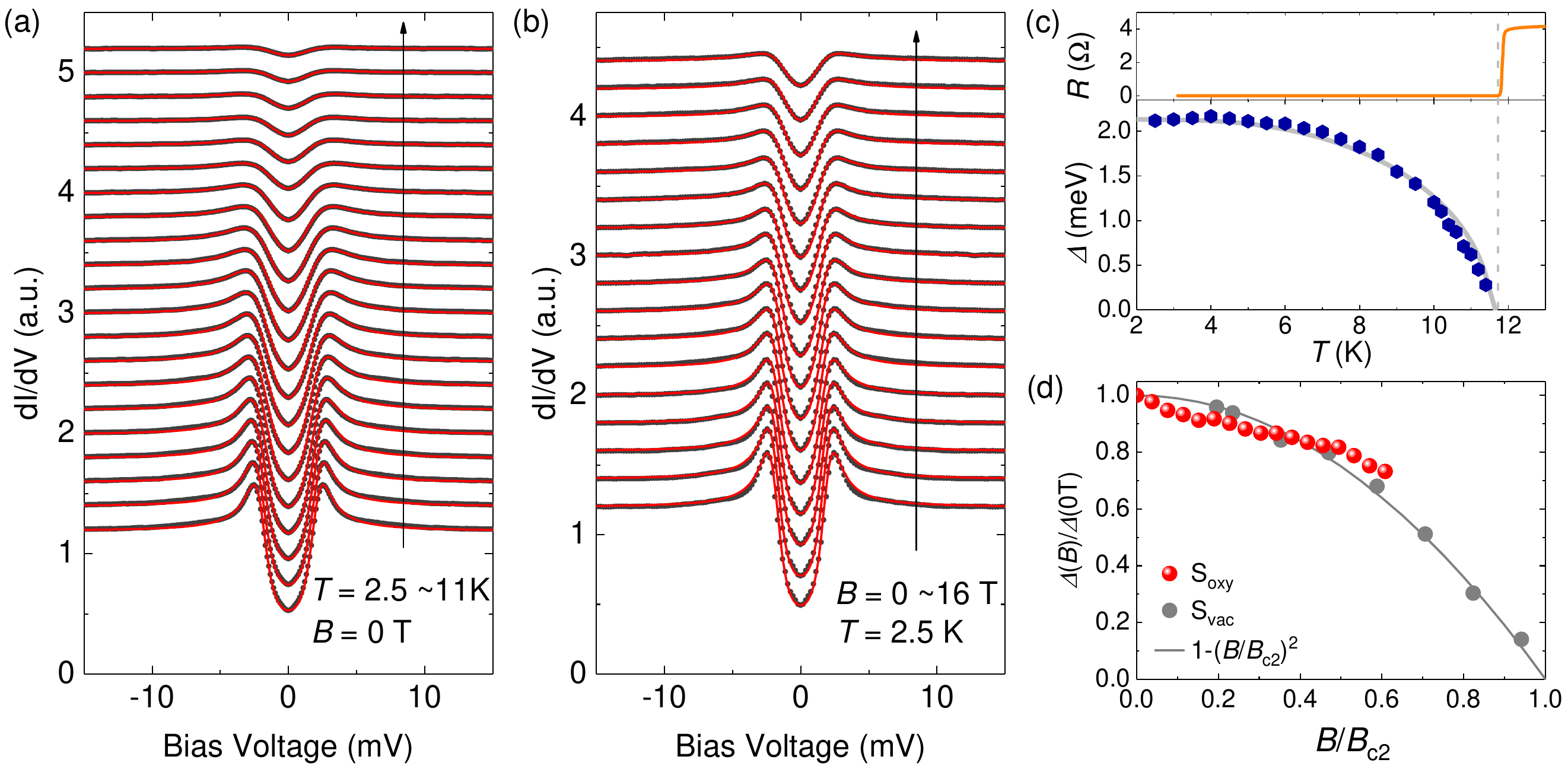}
	\caption{ \textbf{Temperature- and field-dependent tunneling spectra and the superconducting energy gap of S$_\mathrm{oxy}^{\prime}$.} (a)~Normalized differential conductance curves from $2.5$~K to $10$~K with $\Delta T = 0.5~K$ and from $10.2$~K to $11$~K with $\Delta T = 0.2~K$ in zero field. All the data are vertically shifted for clarity. (b)~Normalized differential conductance versus field from $0$~T to $16$~T with $\Delta B = 1~T$ at $2.5$~K. Data in fields are vertically shifted. Both sets of experimental data (black dots) are fitted by the modified BTK model (red lines). (c)~The top panel shows $R(T)$ curve of S$_\mathrm{oxy}^{\prime}$. The temperature-dependent energy gap values (blue hexagons) shown in bottom panel are fitted by the BCS theory (grey line). The vertical dash line indicates that the closing temperature of energy gap is consistent with $T_\mathrm{c}$. (d)~Normalized energy gap versus $B/B_\mathrm{c2}$ (red spheres). The normalized gap values of samples deposited in high vacuum (grey dots) are extracted from our previous work\cite{Jin2015}. The grey line is plotted by the function, $\Delta (B)/\Delta (0~\mathrm{T}) = 1-(B/B_\mathrm{c2})^2$.}
	\label{fig3}
\end{figure}

To further clarify the key factors that lead to anomalies of $B_\mathrm{c2}$, we carried out point-contact tunneling spectroscopy measurements for S$_\mathrm{oxy}^{\prime}$ whose RRR $= 3.4$. Fig.~\ref{fig3}~(a) and (b) display the normalized temperature- and field-dependent tunneling spectra respectively, with $B$ perpendicular to $ab$-plane. The differential conductance spectra clearly show a pair of superconducting coherence peaks, which are still robust against field up to $16$~T at $2.5$~K. All the spectra are fitted well with the framework of Blonder-Tinkham-Klapwijk (BTK) model\cite{Jin2015}. The temperature-dependent superconducting energy gap $\Delta (T)$ obtained from the BTK fits, agrees well with BCS theory and $2\Delta / k_\mathrm{B}T_\mathrm{c} = 4.2$ where $k_\mathrm{B}$ is the Boltzmann constant, as shown in the bottom panel of Fig.~\ref{fig3}~(c). Compared with S$_\mathrm{vac}$, $2\Delta / k_\mathrm{B}T_\mathrm{c}$ of S$_\mathrm{oxy}^{\prime}$ is nearly unchanged\cite{Jin2015,He2017_LTO}, indicating that the electron-phonon coupling remains the same. However, the normalized field-dependent energy gap $\Delta (B)/\Delta (0)$ of S$_\mathrm{oxy}^{\prime}$ versus $B$ deviates from the relation, $\Delta (B)/\Delta (0) = 1-(B/B_\mathrm{c2})^2$, which is discovered in S$_\mathrm{vac}$\cite{Jin2015}, as shown in Fig.~\ref{fig3}~(d). This unusual behavior of $\Delta (B)$ suggests that orbital order is suppressed while spin-orbit scattering might be enhanced in S$_\mathrm{oxy}^{\prime}$, coinciding with our previous work\cite{Jia2018}.

\begin{figure*}[ht!]
	\centering
	\includegraphics[width=\linewidth]{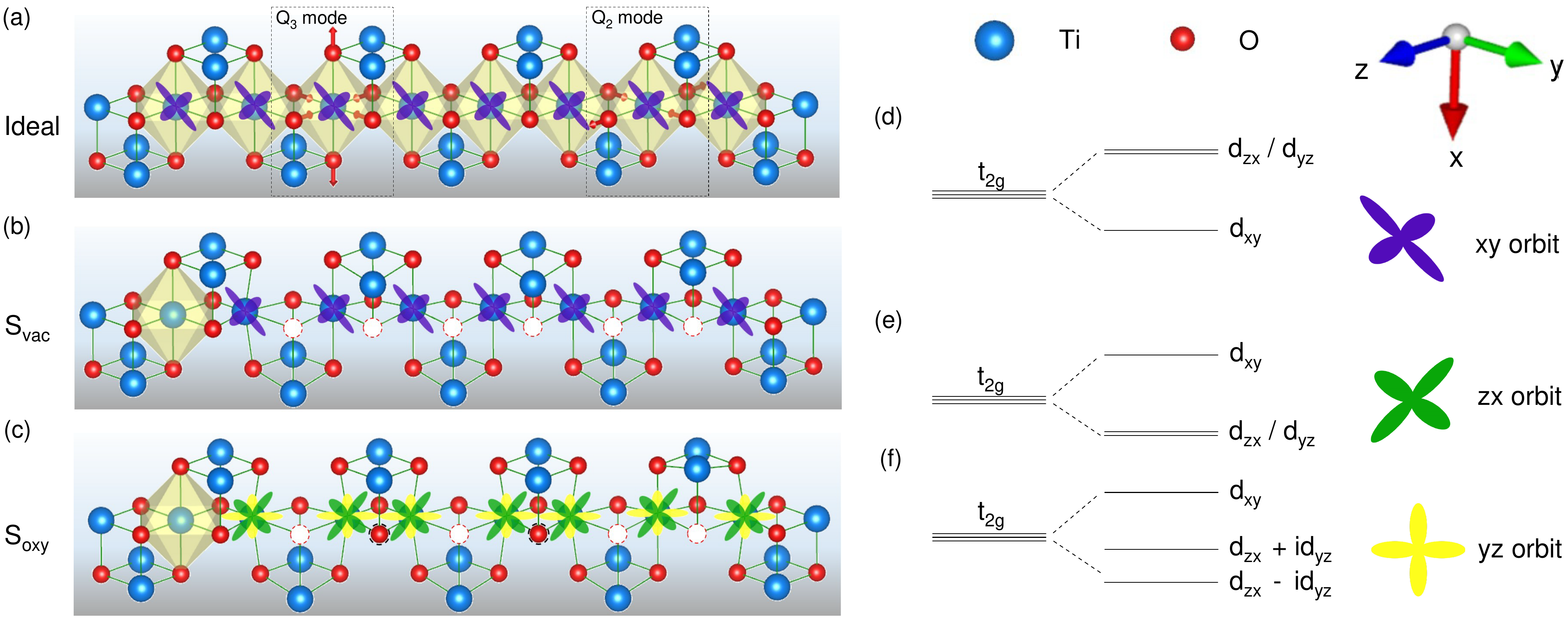}
	\caption{ \textbf{The crystal structures and proposed band splitting of LiTi$_2$O$_{4-\delta}$.} (a)~Octahedron chain in ideal LiTi$_2$O$_{4}$ structure. The red arrows show the phonon vibration modes, i.e. Q$_{2}$ and Q$_{3}$ modes. $d_{xy}$ orbits are shown in the Ti atoms. (b)~Octahedron chain in S$_\mathrm{vac}$. The dotted circles stand for the oxygen vacancies. $d_{xy}$ orbits are shown in the Ti atoms. (c)~Octahedron chain in S$_\mathrm{oxy}$. The oxygen atoms surrounded by dotted circles stand for the filled oxygen vacancies. $d_{zx} + id_{yz}$ and $d_{zx} - id_{yz}$ are shown in the Ti atoms. (d)~Orbital level splitting induced by Jahn-Teller distortion in ideal LiTi$_2$O$_{4}$ structure and S$_\mathrm{vac}$. (e)~Orbital level splitting induced by Jahn-Teller distortion with the existence of oxygen impurities. (f)~Orbital level splitting induced by Jahn-Teller distortion and spin-orbit coupling in S$_\mathrm{oxy}$.}
	\label{fig4}
\end{figure*}

Now let's turn to discuss the mechanisms behind the enhancement of $B_\mathrm{c2}$ for LiTi$_2$O$_{4-\delta}$ films. As mentioned above, $B_\mathrm{c2}$ can be improved by suppressing orbital and/or spin depairing effects. It seems that orbital effect is crucial since $\alpha$, as a key factor to characterize the orbital contribution\cite{Maki_III}, is significantly enhanced in S$_\mathrm{oxy}$. According to the WHH theory, $\alpha = 3 \hbar / (4 E_\mathrm{F} \tau)$, where $\hbar$, $E_\mathrm{F}$ and $\tau$ are the reduced Planck constant, Fermi energy and relaxation time respectively. We can obtain $\tau = 1.4 \times 10^{-14}$~s for S$_\mathrm{vac}$, which is self-consistent with the result derived from electrical transport, i.e. $\tau = l / v_{F} = 1.3 \times 10^{-14}$~s, where $l$ is mean free path and $v_\mathrm{F}$ is Fermi velocity\cite{Jin2015}. However, the excellent fitting of S$_\mathrm{oxy}$ provides an underestimation of $\tau$ ($= 5 \times 10^{-16}$~s) if we assume the Fermi energy is constant, since such $\tau$ is even smaller than the lower limit $\tau_\mathrm{min}$ ($= c / v_\mathrm{F} = 6.1 \times 10^{-15}$~s, where $c$ is the out-of-plane lattice constant), restricted by the Mott-Ioffe-Regel limit\cite{Hussey2004,Jia2018}. Such confliction can be reconciled in case of a reduced $E_\mathrm{F}^{oxy} \sim E_\mathrm{F}^{vac} / 8$, where the label $oxy$ and $vac$ stand for S$_\mathrm{oxy}$ and S$_\mathrm{vac}$, respectively. It is surprising that tiny amount of oxygen defects can induce such remarkable change in Fermi energy unless the occurrence of Fermi surface reconstruction or Lifshitz transition and formation of small Fermi pockets, like in cuprates\cite{Matsui2007}. We expect that such variation of Fermi energy should have influence on density of states thereby tuning $T_\mathrm{c}$ and $\Delta$, which is not observed in our work. Besides, some effects beyond the WHH theory such as dimensionality and strong electron-phonon coupling can also be easily ruled out since the thickness of samples ($> 70$~nm) is much larger than the lattice constant and $2 \Delta / (k_\mathrm{B} T_\mathrm{c})$ is almost unchanged for all samples.

Alternatively, the anomalies of $B_\mathrm{c2}$ may be attributed to some factors that suppress spin flip, such as SOC and SOS. Zeeman SOC and Rashba SOC are not the case of LiTi$_2$O$_{4-\delta}$ because of the centrosymmetric lattice structure and isotropic $B_\mathrm{c2}$\cite{Lu2015,Sohn2018}. SOS is a candidate to understand the enhancement of $B_\mathrm{c2}$. Nevertheless, the values of $\lambda_\mathrm{so}$ derived from the WHH fitting are too small to induce such large enhancement compared with other systems\cite{Fischer1978}. It should be emphasized that $s$-wave scattering is assumed in the WHH theory, while other factors like the details of spin-orbit interaction are ignored\cite{Werthamer1966,Rieck1991}, which may also play an essential role in the anomalies of $B_\mathrm{c2}$. In order to clarify this issue, we consider the connection between the crystal structures and spin-orbit interaction in three configurations, i.e. ideal LiTi$_2$O$_{4}$ structure (see Fig.~\ref{fig4}~(a)), existence of long range ordered oxygen vacancies (S$_\mathrm{vac}$, see Fig.~\ref{fig4}~(b)) and the case where these vacancies are partially filled by oxygen atoms (S$_\mathrm{oxy}$, see Fig.~\ref{fig4}~(c)). Since it is impossible to take into account oxygen impurities with random distribution in band structure calculations, we would like to discuss it phenomenologically. For the ideal LiTi$_2$O$_{4}$ structure, Jahn-Teller distortions with Q$_{2}$ and Q$_{3}$ modes exist homogeneously\cite{Longuethiggins1958}. In this case, the compression of TiO$_{6}$ octahedron dominates, and $t_{2g}$ orbits split into higher two-fold degenerate $d_{zx}$/$d_{yz}$ and lower non-degenerate $d_{xy}$\cite{Khomskii2005} (see Fig.~\ref{fig4}~(d)). If long range ordered oxygen vacancies exist (S$_\mathrm{vac}$), such distortion will be stabilized, and orbital-related state is expected to form, which is indeed observed in S$_\mathrm{vac}$ \cite{Jin2015,Jia2018}. Here on the single octahedron the orbital moment of the conduction electron quenches out.

However, spin-orbit interaction should be taken into account even if only one oxygen vacancy is filled since the gradient of local electrical potential $V$ cannot be neglected. In general, the Hamiltonian term of SOC can be expressed as
\begin{equation}\label{eq1}
  H_\mathrm{soc} = \lambda_\mathrm{soc} (\nabla V \times \textbf{k}) \cdot \boldsymbol{\sigma} ~,
\end{equation}
where $\lambda_\mathrm{soc}$ is the spin-orbit coupling constant, $\textbf{k}$ is the momentum of the conduction electron, $\boldsymbol{\sigma}$ is the Pauli matrix. It is well known that $\lambda_\mathrm{soc}$ strongly increases with the atomic number Z, i.e. $\lambda_\mathrm{soc}\sim \mathrm{Z}^4$\cite{Watkinsi2003}. In this case, the energy scale of SOC is comparable to that of Jahn-Teller distortions\cite{Watkinsi2003,Haverkort2005}. Therefore, the SOC competes with the Jahn-Teller effect and leads to a reversal of orbital splitting, i.e. higher non-degenerate $d_{xy}$ and lower two-fold degenerate $d_{zx}$/$d_{yz}$\cite{Bersuker2006}(see Fig.~\ref{fig4}~(e)). Typically, the SOC overbears the types of Jahn-Teller distortion, and further splits two-fold degenerate $d_{zx}$/$d_{yz}$ into $d_{zx} + id_{yz}$ and $d_{zx} - id_{yz}$ (see Fig.~\ref{fig4}~(f)), consequently generating an orbital moment quantum number $l=\pm1$. As a result, when an external magnetic field $B$ is applied, the effective field $B_\mathrm{eff}$ is given by
\begin{equation}\label{eq1}
  B_\mathrm{eff} = B + \langle B_\mathrm{soc}\rangle ~,
\end{equation}
where $\langle B_\mathrm{soc}\rangle$ is the average value of the generated SOC field. Since $\langle B_\mathrm{soc}\rangle$ is negative due to the opposite orientations of orbital moment and spin\cite{Mizokawa1996,Meijer1999}, the external field is weakened and thus the $B_\mathrm{c2}$ is enhanced. The similar situation with the orbital moment orientations has been discussed in another Ti oxide compound LaTiO$_{3}$, which has a similar ordering of orbital energies\cite{Mizokawa1996,Meijer1999}. Besides, momentum-dependent spin-orbit scattering may also contribute to such enhancement of $B_\mathrm{c2}$\cite{Neuringer1966}. As shown long ago by Boiko and Rashba, it is very important for magnetic susceptibility and other magnetic properties of materials with spin-orbit interaction\cite{Rashba1960}.


Other anomalies can be well understood in this framework. Firstly, due to the nonselective distribution of oxygen impurities, such interaction is isotropic in average, leading to an isotropic $B_\mathrm{c2}$. Secondly, such spin-orbit interaction has no effect on the thermodynamics of a superconductor, especially $T_\mathrm{c}$, pointed out by Gor$^\prime$kov and Rusinov\cite{Gorkov1964}, which is consistent with our results. Thirdly, the deviation from $\Delta (B) \sim -B^2$ in S$_\mathrm{oxy}$ indicates the suppression of orbital order via the randomly oriented and distributed orbital moments, therewith the enhanced spin-orbit interaction.


Overall, by a systematic transport and point-contact tunneling spectroscopy measurements of the LiTi$_2$O$_{4-\delta}$ films we find an astonishing enhancement of the $B_\mathrm{c2}$, that breaks the Pauli limit and meanwhile remains isotropic. Such anomalies are rarely observed in superconductors with cubic structure, yet the breaking of the Pauli limit is frequently reported in (quasi) two dimensional non-centrosymmetric superconducting materials but with an anisotropic $B_\mathrm{c2}$. Moreover, oxygen impurities, giving rise to higher residual resistivity, do not make obvious influence on $T_\mathrm{c}$ and $\Delta$. In combination with our previous work on this system\cite{Jin2015,Jia2018}, we conclude that the three dimensional spin-orbit interaction induced by oxygen impurities is essential to the significant isotropic enhancement of $B_\mathrm{c2}$. The effects of oxygen impurities could be general in oxides, thus providing us new clues to generate other exotic phenomena such as Ising superconductivity\cite{Lu2015,Zhou2016}, topological nontrivial state due to band inversion\cite{Kusmartsev1985,Dmytro2010,Galitski2013}, vortex phase transition from liquid to solid\cite{Rykov1999}. In addition, our achievements pave a promising path to optimize practical superconductors with higher upper critical filed.

We thank X. Zhang, X. Y. Jiang, D. Li, J. S. Zhang, A. Kusmartsev for fruitful discussions and E. I. Rashba for useful comments and indicating the Ref.\cite{Rashba1960}. This work was supported by the National Key Basic Research Program of China (2015CB921000, 2017YFA0303003, 2017YFA0302902 and 2018YFB0704102), the National Natural Science Foundation of China (11674374 and 11804378), the Key Research Program of Frontier Sciences, CAS (QYZDB-SSW-SLH008 and QYZDY-SSW-SLH001), the Key Research Program of the Chinese Academy of Sciences (XDPB01). The work of F.V.K. was supported by the Government of the Russian Federation through the ITMO Professorship Program. A portion of this work was performed on the Steady High Magnetic Field Facilities, High Magnetic Field Laboratory, CAS.

\nocite{*}
\bibliography{ref} 

\begin{thebibliography}{39}%
\makeatletter
\providecommand \@ifxundefined [1]{%
 \@ifx{#1\undefined}
}%
\providecommand \@ifnum [1]{%
 \ifnum #1\expandafter \@firstoftwo
 \else \expandafter \@secondoftwo
 \fi
}%
\providecommand \@ifx [1]{%
 \ifx #1\expandafter \@firstoftwo
 \else \expandafter \@secondoftwo
 \fi
}%
\providecommand \natexlab [1]{#1}%
\providecommand \enquote  [1]{``#1''}%
\providecommand \bibnamefont  [1]{#1}%
\providecommand \bibfnamefont [1]{#1}%
\providecommand \citenamefont [1]{#1}%
\providecommand \href@noop [0]{\@secondoftwo}%
\providecommand \href [0]{\begingroup \@sanitize@url \@href}%
\providecommand \@href[1]{\@@startlink{#1}\@@href}%
\providecommand \@@href[1]{\endgroup#1\@@endlink}%
\providecommand \@sanitize@url [0]{\catcode `\\12\catcode `\$12\catcode
  `\&12\catcode `\#12\catcode `\^12\catcode `\_12\catcode `\%12\relax}%
\providecommand \@@startlink[1]{}%
\providecommand \@@endlink[0]{}%
\providecommand \url  [0]{\begingroup\@sanitize@url \@url }%
\providecommand \@url [1]{\endgroup\@href {#1}{\urlprefix }}%
\providecommand \urlprefix  [0]{URL }%
\providecommand \Eprint [0]{\href }%
\providecommand \doibase [0]{http://dx.doi.org/}%
\providecommand \selectlanguage [0]{\@gobble}%
\providecommand \bibinfo  [0]{\@secondoftwo}%
\providecommand \bibfield  [0]{\@secondoftwo}%
\providecommand \translation [1]{[#1]}%
\providecommand \BibitemOpen [0]{}%
\providecommand \bibitemStop [0]{}%
\providecommand \bibitemNoStop [0]{.\EOS\space}%
\providecommand \EOS [0]{\spacefactor3000\relax}%
\providecommand \BibitemShut  [1]{\csname bibitem#1\endcsname}%
\let\auto@bib@innerbib\@empty
\bibitem [{\citenamefont {Saint-James}\ \emph {et~al.}(1969)\citenamefont
  {Saint-James}, \citenamefont {Sarma},\ and\ \citenamefont
  {Thomas}}]{James1969}%
  \BibitemOpen
  \bibfield  {author} {\bibinfo {author} {\bibfnamefont {D.}~\bibnamefont
  {Saint-James}}, \bibinfo {author} {\bibfnamefont {G.}~\bibnamefont {Sarma}},
  \ and\ \bibinfo {author} {\bibfnamefont {E.~J.}\ \bibnamefont {Thomas}},\
  }\href@noop {} {\emph {\bibinfo {title} {Type II Superconductivity}}}\
  (\bibinfo  {publisher} {Pergamon Press},\ \bibinfo {address} {Oxford},\
  \bibinfo {year} {1969})\BibitemShut {NoStop}%
\bibitem [{\citenamefont {Sohn}\ \emph {et~al.}(2018)\citenamefont {Sohn},
  \citenamefont {Xi}, \citenamefont {He}, \citenamefont {Jiang}, \citenamefont
  {Wang}, \citenamefont {Kang}, \citenamefont {Park}, \citenamefont {Berger},
  \citenamefont {Forr\'o}, \citenamefont {Law}, \citenamefont {Shan},\ and\
  \citenamefont {Mak}}]{Sohn2018}%
  \BibitemOpen
  \bibfield  {author} {\bibinfo {author} {\bibfnamefont {E.}~\bibnamefont
  {Sohn}}, \bibinfo {author} {\bibfnamefont {X.}~\bibnamefont {Xi}}, \bibinfo
  {author} {\bibfnamefont {W.~Y.}\ \bibnamefont {He}}, \bibinfo {author}
  {\bibfnamefont {S.}~\bibnamefont {Jiang}}, \bibinfo {author} {\bibfnamefont
  {Z.}~\bibnamefont {Wang}}, \bibinfo {author} {\bibfnamefont {K.}~\bibnamefont
  {Kang}}, \bibinfo {author} {\bibfnamefont {J.~H.}\ \bibnamefont {Park}},
  \bibinfo {author} {\bibfnamefont {H.}~\bibnamefont {Berger}}, \bibinfo
  {author} {\bibfnamefont {L.}~\bibnamefont {Forr\'o}}, \bibinfo {author}
  {\bibfnamefont {K.~T.}\ \bibnamefont {Law}}, \bibinfo {author} {\bibfnamefont
  {J.}~\bibnamefont {Shan}}, \ and\ \bibinfo {author} {\bibfnamefont {K.~F.}\
  \bibnamefont {Mak}},\ }\href {\doibase 10.1038/s41563-018-0061-1} {\bibfield
  {journal} {\bibinfo  {journal} {Nat. Mater.}\ }\textbf {\bibinfo {volume}
  {17}},\ \bibinfo {pages} {504} (\bibinfo {year} {2018})}\BibitemShut
  {NoStop}%
\bibitem [{\citenamefont {Holczer}\ \emph {et~al.}(1991)\citenamefont
  {Holczer}, \citenamefont {Klein}, \citenamefont {Gr\"uner}, \citenamefont
  {Thompson}, \citenamefont {Diederich},\ and\ \citenamefont
  {Whetten}}]{Holczer1991}%
  \BibitemOpen
  \bibfield  {author} {\bibinfo {author} {\bibfnamefont {K.}~\bibnamefont
  {Holczer}}, \bibinfo {author} {\bibfnamefont {O.}~\bibnamefont {Klein}},
  \bibinfo {author} {\bibfnamefont {G.}~\bibnamefont {Gr\"uner}}, \bibinfo
  {author} {\bibfnamefont {J.~D.}\ \bibnamefont {Thompson}}, \bibinfo {author}
  {\bibfnamefont {F.}~\bibnamefont {Diederich}}, \ and\ \bibinfo {author}
  {\bibfnamefont {R.~L.}\ \bibnamefont {Whetten}},\ }\href {\doibase
  10.1103/PhysRevLett.67.271} {\bibfield  {journal} {\bibinfo  {journal} {Phys.
  Rev. Lett.}\ }\textbf {\bibinfo {volume} {67}},\ \bibinfo {pages} {271}
  (\bibinfo {year} {1991})}\BibitemShut {NoStop}%
\bibitem [{\citenamefont {Fischer}(1978)}]{Fischer1978}%
  \BibitemOpen
  \bibfield  {author} {\bibinfo {author} {\bibfnamefont {{\O}.}~\bibnamefont
  {Fischer}},\ }\href {\doibase 10.1007/Bf00931416} {\bibfield  {journal}
  {\bibinfo  {journal} {Appl. Phys.}\ }\textbf {\bibinfo {volume} {16}},\
  \bibinfo {pages} {1} (\bibinfo {year} {1978})}\BibitemShut {NoStop}%
\bibitem [{\citenamefont {Orlando}\ \emph {et~al.}(1979)\citenamefont
  {Orlando}, \citenamefont {McNiff}, \citenamefont {Foner},\ and\ \citenamefont
  {Beasley}}]{Orlando1979}%
  \BibitemOpen
  \bibfield  {author} {\bibinfo {author} {\bibfnamefont {T.~P.}\ \bibnamefont
  {Orlando}}, \bibinfo {author} {\bibfnamefont {E.~J.}\ \bibnamefont {McNiff}},
  \bibinfo {author} {\bibfnamefont {S.}~\bibnamefont {Foner}}, \ and\ \bibinfo
  {author} {\bibfnamefont {M.~R.}\ \bibnamefont {Beasley}},\ }\href {\doibase
  10.1103/PhysRevB.19.4545} {\bibfield  {journal} {\bibinfo  {journal} {Phys.
  Rev. B}\ }\textbf {\bibinfo {volume} {19}},\ \bibinfo {pages} {4545}
  (\bibinfo {year} {1979})}\BibitemShut {NoStop}%
\bibitem [{\citenamefont {Adams}\ \emph {et~al.}(1998)\citenamefont {Adams},
  \citenamefont {Herron},\ and\ \citenamefont {Meletis}}]{Adams1998}%
  \BibitemOpen
  \bibfield  {author} {\bibinfo {author} {\bibfnamefont {P.~W.}\ \bibnamefont
  {Adams}}, \bibinfo {author} {\bibfnamefont {P.}~\bibnamefont {Herron}}, \
  and\ \bibinfo {author} {\bibfnamefont {E.~I.}\ \bibnamefont {Meletis}},\
  }\href {\doibase 10.1103/PhysRevB.58.R2952} {\bibfield  {journal} {\bibinfo
  {journal} {Phys. Rev. B}\ }\textbf {\bibinfo {volume} {58}},\ \bibinfo
  {pages} {R2952} (\bibinfo {year} {1998})}\BibitemShut {NoStop}%
\bibitem [{\citenamefont {Tedrow}\ and\ \citenamefont
  {Meservey}(1982)}]{Tedrow1982}%
  \BibitemOpen
  \bibfield  {author} {\bibinfo {author} {\bibfnamefont {P.~M.}\ \bibnamefont
  {Tedrow}}\ and\ \bibinfo {author} {\bibfnamefont {R.}~\bibnamefont
  {Meservey}},\ }\href {\doibase 10.1103/PhysRevB.25.171} {\bibfield  {journal}
  {\bibinfo  {journal} {Phys. Rev. B}\ }\textbf {\bibinfo {volume} {25}},\
  \bibinfo {pages} {171} (\bibinfo {year} {1982})}\BibitemShut {NoStop}%
\bibitem [{\citenamefont {Clogston}(1962)}]{Clogston1962}%
  \BibitemOpen
  \bibfield  {author} {\bibinfo {author} {\bibfnamefont {A.~M.}\ \bibnamefont
  {Clogston}},\ }\href {\doibase 10.1103/PhysRevLett.9.266} {\bibfield
  {journal} {\bibinfo  {journal} {Phys. Rev. Lett.}\ }\textbf {\bibinfo
  {volume} {9}},\ \bibinfo {pages} {266} (\bibinfo {year} {1962})}\BibitemShut
  {NoStop}%
\bibitem [{\citenamefont {Chandrasekhar}(1962)}]{Chandrasekhar1962}%
  \BibitemOpen
  \bibfield  {author} {\bibinfo {author} {\bibfnamefont {B.~S.}\ \bibnamefont
  {Chandrasekhar}},\ }\href {\doibase 10.1063/1.1777362} {\bibfield  {journal}
  {\bibinfo  {journal} {Appl. Phys. Lett.}\ }\textbf {\bibinfo {volume} {1}},\
  \bibinfo {pages} {7} (\bibinfo {year} {1962})}\BibitemShut {NoStop}%
\bibitem [{\citenamefont {Matsuda}\ and\ \citenamefont
  {Shimahara}(2007)}]{Matsuda2007}%
  \BibitemOpen
  \bibfield  {author} {\bibinfo {author} {\bibfnamefont {Y.}~\bibnamefont
  {Matsuda}}\ and\ \bibinfo {author} {\bibfnamefont {H.}~\bibnamefont
  {Shimahara}},\ }\href {\doibase 10.1143/Jpsj.76.051005} {\bibfield  {journal}
  {\bibinfo  {journal} {J. Phys. Soc. Jpn.}\ }\textbf {\bibinfo {volume}
  {76}},\ \bibinfo {pages} {051005} (\bibinfo {year} {2007})}\BibitemShut
  {NoStop}%
\bibitem [{\citenamefont {Huy}\ \emph {et~al.}(2007)\citenamefont {Huy},
  \citenamefont {Gasparini}, \citenamefont {de~Nijs}, \citenamefont {Huang},
  \citenamefont {Klaasse}, \citenamefont {Gortenmulder}, \citenamefont
  {de~Visser}, \citenamefont {Hamann}, \citenamefont {G\"orlach},\ and\
  \citenamefont {L\"ohneysen}}]{Huy2007}%
  \BibitemOpen
  \bibfield  {author} {\bibinfo {author} {\bibfnamefont {N.~T.}\ \bibnamefont
  {Huy}}, \bibinfo {author} {\bibfnamefont {A.}~\bibnamefont {Gasparini}},
  \bibinfo {author} {\bibfnamefont {D.~E.}\ \bibnamefont {de~Nijs}}, \bibinfo
  {author} {\bibfnamefont {Y.}~\bibnamefont {Huang}}, \bibinfo {author}
  {\bibfnamefont {J.~C.~P.}\ \bibnamefont {Klaasse}}, \bibinfo {author}
  {\bibfnamefont {T.}~\bibnamefont {Gortenmulder}}, \bibinfo {author}
  {\bibfnamefont {A.}~\bibnamefont {de~Visser}}, \bibinfo {author}
  {\bibfnamefont {A.}~\bibnamefont {Hamann}}, \bibinfo {author} {\bibfnamefont
  {T.}~\bibnamefont {G\"orlach}}, \ and\ \bibinfo {author} {\bibfnamefont
  {H.~v.}\ \bibnamefont {L\"ohneysen}},\ }\href {\doibase
  10.1103/PhysRevLett.99.067006} {\bibfield  {journal} {\bibinfo  {journal}
  {Phys. Rev. Lett.}\ }\textbf {\bibinfo {volume} {99}},\ \bibinfo {pages}
  {067006} (\bibinfo {year} {2007})}\BibitemShut {NoStop}%
\bibitem [{\citenamefont {Aoki}\ and\ \citenamefont
  {Flouquet}(2012)}]{Aoki2012}%
  \BibitemOpen
  \bibfield  {author} {\bibinfo {author} {\bibfnamefont {D.}~\bibnamefont
  {Aoki}}\ and\ \bibinfo {author} {\bibfnamefont {J.}~\bibnamefont
  {Flouquet}},\ }\href {\doibase 01100310.1143/Jpsj.81.011003} {\bibfield
  {journal} {\bibinfo  {journal} {J. Phys. Soc. Jpn.}\ }\textbf {\bibinfo
  {volume} {81}},\ \bibinfo {pages} {011003} (\bibinfo {year}
  {2012})}\BibitemShut {NoStop}%
\bibitem [{\citenamefont {Herranz}\ \emph {et~al.}(2015)\citenamefont
  {Herranz}, \citenamefont {Singh}, \citenamefont {Bergeal}, \citenamefont
  {Jouan}, \citenamefont {Lesueur}, \citenamefont {Gazquez}, \citenamefont
  {Varela}, \citenamefont {Scigaj}, \citenamefont {Dix}, \citenamefont
  {Sanchez},\ and\ \citenamefont {Fontcuberta}}]{Herranz2015}%
  \BibitemOpen
  \bibfield  {author} {\bibinfo {author} {\bibfnamefont {G.}~\bibnamefont
  {Herranz}}, \bibinfo {author} {\bibfnamefont {G.}~\bibnamefont {Singh}},
  \bibinfo {author} {\bibfnamefont {N.}~\bibnamefont {Bergeal}}, \bibinfo
  {author} {\bibfnamefont {A.}~\bibnamefont {Jouan}}, \bibinfo {author}
  {\bibfnamefont {J.}~\bibnamefont {Lesueur}}, \bibinfo {author} {\bibfnamefont
  {J.}~\bibnamefont {Gazquez}}, \bibinfo {author} {\bibfnamefont
  {M.}~\bibnamefont {Varela}}, \bibinfo {author} {\bibfnamefont
  {M.}~\bibnamefont {Scigaj}}, \bibinfo {author} {\bibfnamefont
  {N.}~\bibnamefont {Dix}}, \bibinfo {author} {\bibfnamefont {F.}~\bibnamefont
  {Sanchez}}, \ and\ \bibinfo {author} {\bibfnamefont {J.}~\bibnamefont
  {Fontcuberta}},\ }\href {\doibase 10.1038/Ncomms7028} {\bibfield  {journal}
  {\bibinfo  {journal} {Nat. Commun.}\ }\textbf {\bibinfo {volume} {6}},\
  \bibinfo {pages} {6028} (\bibinfo {year} {2015})}\BibitemShut {NoStop}%
\bibitem [{\citenamefont {Lu}\ \emph {et~al.}(2015)\citenamefont {Lu},
  \citenamefont {Zheliuk}, \citenamefont {Leermakers}, \citenamefont {Yuan},
  \citenamefont {Zeitler}, \citenamefont {Law},\ and\ \citenamefont
  {Ye}}]{Lu2015}%
  \BibitemOpen
  \bibfield  {author} {\bibinfo {author} {\bibfnamefont {J.~M.}\ \bibnamefont
  {Lu}}, \bibinfo {author} {\bibfnamefont {O.}~\bibnamefont {Zheliuk}},
  \bibinfo {author} {\bibfnamefont {I.}~\bibnamefont {Leermakers}}, \bibinfo
  {author} {\bibfnamefont {N.~F.~Q.}\ \bibnamefont {Yuan}}, \bibinfo {author}
  {\bibfnamefont {U.}~\bibnamefont {Zeitler}}, \bibinfo {author} {\bibfnamefont
  {K.~T.}\ \bibnamefont {Law}}, \ and\ \bibinfo {author} {\bibfnamefont
  {J.~T.}\ \bibnamefont {Ye}},\ }\href {\doibase 10.1126/science.aab2277}
  {\bibfield  {journal} {\bibinfo  {journal} {Science}\ }\textbf {\bibinfo
  {volume} {350}},\ \bibinfo {pages} {1353} (\bibinfo {year}
  {2015})}\BibitemShut {NoStop}%
\bibitem [{\citenamefont {Maki}\ and\ \citenamefont {Tsuneto}(1964)}]{Maki_I}%
  \BibitemOpen
  \bibfield  {author} {\bibinfo {author} {\bibfnamefont {K.}~\bibnamefont
  {Maki}}\ and\ \bibinfo {author} {\bibfnamefont {T.}~\bibnamefont {Tsuneto}},\
  }\href {\doibase 10.1143/PTP.31.945} {\bibfield  {journal} {\bibinfo
  {journal} {Prog. Theor. Phys.}\ }\textbf {\bibinfo {volume} {31}},\ \bibinfo
  {pages} {945} (\bibinfo {year} {1964})}\BibitemShut {NoStop}%
\bibitem [{\citenamefont {Jin}\ \emph {et~al.}(2015)\citenamefont {Jin},
  \citenamefont {He}, \citenamefont {Zhang}, \citenamefont {Maruyama},
  \citenamefont {Yasui}, \citenamefont {Suchoski}, \citenamefont {Shin},
  \citenamefont {Jiang}, \citenamefont {Yu}, \citenamefont {Yuan},
  \citenamefont {Shan}, \citenamefont {Kusmartsev}, \citenamefont {Greene},\
  and\ \citenamefont {Takeuchi}}]{Jin2015}%
  \BibitemOpen
  \bibfield  {author} {\bibinfo {author} {\bibfnamefont {K.}~\bibnamefont
  {Jin}}, \bibinfo {author} {\bibfnamefont {G.}~\bibnamefont {He}}, \bibinfo
  {author} {\bibfnamefont {X.}~\bibnamefont {Zhang}}, \bibinfo {author}
  {\bibfnamefont {S.}~\bibnamefont {Maruyama}}, \bibinfo {author}
  {\bibfnamefont {S.}~\bibnamefont {Yasui}}, \bibinfo {author} {\bibfnamefont
  {R.}~\bibnamefont {Suchoski}}, \bibinfo {author} {\bibfnamefont
  {J.}~\bibnamefont {Shin}}, \bibinfo {author} {\bibfnamefont {Y.}~\bibnamefont
  {Jiang}}, \bibinfo {author} {\bibfnamefont {H.~S.}\ \bibnamefont {Yu}},
  \bibinfo {author} {\bibfnamefont {J.}~\bibnamefont {Yuan}}, \bibinfo {author}
  {\bibfnamefont {L.}~\bibnamefont {Shan}}, \bibinfo {author} {\bibfnamefont
  {F.~V.}\ \bibnamefont {Kusmartsev}}, \bibinfo {author} {\bibfnamefont
  {R.~L.}\ \bibnamefont {Greene}}, \ and\ \bibinfo {author} {\bibfnamefont
  {I.}~\bibnamefont {Takeuchi}},\ }\href {\doibase 10.1038/ncomms8183}
  {\bibfield  {journal} {\bibinfo  {journal} {Nat. Commun.}\ }\textbf {\bibinfo
  {volume} {6}},\ \bibinfo {pages} {7183} (\bibinfo {year} {2015})}\BibitemShut
  {NoStop}%
\bibitem [{\citenamefont {Satpathy}\ and\ \citenamefont
  {Martin}(1987)}]{Satpathy1987}%
  \BibitemOpen
  \bibfield  {author} {\bibinfo {author} {\bibfnamefont {S.}~\bibnamefont
  {Satpathy}}\ and\ \bibinfo {author} {\bibfnamefont {R.~M.}\ \bibnamefont
  {Martin}},\ }\href {\doibase 10.1103/PhysRevB.36.7269} {\bibfield  {journal}
  {\bibinfo  {journal} {Phys. Rev. B}\ }\textbf {\bibinfo {volume} {36}},\
  \bibinfo {pages} {7269} (\bibinfo {year} {1987})}\BibitemShut {NoStop}%
\bibitem [{\citenamefont {He}\ \emph {et~al.}(2017)\citenamefont {He},
  \citenamefont {Jia}, \citenamefont {Hou}, \citenamefont {Wei}, \citenamefont
  {Xie}, \citenamefont {Yang}, \citenamefont {Shi}, \citenamefont {Yuan},
  \citenamefont {Shan}, \citenamefont {Zhu}, \citenamefont {Li}, \citenamefont
  {Gu}, \citenamefont {Liu}, \citenamefont {Xiang},\ and\ \citenamefont
  {Jin}}]{He2017_LTO}%
  \BibitemOpen
  \bibfield  {author} {\bibinfo {author} {\bibfnamefont {G.}~\bibnamefont
  {He}}, \bibinfo {author} {\bibfnamefont {Y.}~\bibnamefont {Jia}}, \bibinfo
  {author} {\bibfnamefont {X.}~\bibnamefont {Hou}}, \bibinfo {author}
  {\bibfnamefont {Z.}~\bibnamefont {Wei}}, \bibinfo {author} {\bibfnamefont
  {H.}~\bibnamefont {Xie}}, \bibinfo {author} {\bibfnamefont {Z.}~\bibnamefont
  {Yang}}, \bibinfo {author} {\bibfnamefont {J.}~\bibnamefont {Shi}}, \bibinfo
  {author} {\bibfnamefont {J.}~\bibnamefont {Yuan}}, \bibinfo {author}
  {\bibfnamefont {L.}~\bibnamefont {Shan}}, \bibinfo {author} {\bibfnamefont
  {B.}~\bibnamefont {Zhu}}, \bibinfo {author} {\bibfnamefont {H.}~\bibnamefont
  {Li}}, \bibinfo {author} {\bibfnamefont {L.}~\bibnamefont {Gu}}, \bibinfo
  {author} {\bibfnamefont {K.}~\bibnamefont {Liu}}, \bibinfo {author}
  {\bibfnamefont {T.}~\bibnamefont {Xiang}}, \ and\ \bibinfo {author}
  {\bibfnamefont {K.}~\bibnamefont {Jin}},\ }\href {\doibase
  10.1103/PhysRevB.95.054510} {\bibfield  {journal} {\bibinfo  {journal} {Phys.
  Rev. B}\ }\textbf {\bibinfo {volume} {95}},\ \bibinfo {pages} {054510}
  (\bibinfo {year} {2017})}\BibitemShut {NoStop}%
\bibitem [{\citenamefont {Jia}\ \emph {et~al.}(2018)\citenamefont {Jia},
  \citenamefont {He}, \citenamefont {Hu}, \citenamefont {Yang}, \citenamefont
  {Yang}, \citenamefont {Yu}, \citenamefont {Zhang}, \citenamefont {Shi},
  \citenamefont {Lin}, \citenamefont {Yuan}, \citenamefont {Zhu}, \citenamefont
  {Gu}, \citenamefont {Li},\ and\ \citenamefont {Jin}}]{Jia2018}%
  \BibitemOpen
  \bibfield  {author} {\bibinfo {author} {\bibfnamefont {Y.}~\bibnamefont
  {Jia}}, \bibinfo {author} {\bibfnamefont {G.}~\bibnamefont {He}}, \bibinfo
  {author} {\bibfnamefont {W.}~\bibnamefont {Hu}}, \bibinfo {author}
  {\bibfnamefont {H.}~\bibnamefont {Yang}}, \bibinfo {author} {\bibfnamefont
  {Z.}~\bibnamefont {Yang}}, \bibinfo {author} {\bibfnamefont {H.}~\bibnamefont
  {Yu}}, \bibinfo {author} {\bibfnamefont {Q.}~\bibnamefont {Zhang}}, \bibinfo
  {author} {\bibfnamefont {J.}~\bibnamefont {Shi}}, \bibinfo {author}
  {\bibfnamefont {Z.}~\bibnamefont {Lin}}, \bibinfo {author} {\bibfnamefont
  {J.}~\bibnamefont {Yuan}}, \bibinfo {author} {\bibfnamefont {B.}~\bibnamefont
  {Zhu}}, \bibinfo {author} {\bibfnamefont {L.}~\bibnamefont {Gu}}, \bibinfo
  {author} {\bibfnamefont {H.}~\bibnamefont {Li}}, \ and\ \bibinfo {author}
  {\bibfnamefont {K.}~\bibnamefont {Jin}},\ }\href {\doibase
  10.1038/s41598-018-22393-8} {\bibfield  {journal} {\bibinfo  {journal} {Sci.
  Rep.}\ }\textbf {\bibinfo {volume} {8}},\ \bibinfo {pages} {3995} (\bibinfo
  {year} {2018})}\BibitemShut {NoStop}%
\bibitem [{\citenamefont {Werthamer}\ \emph {et~al.}(1966)\citenamefont
  {Werthamer}, \citenamefont {Helfand},\ and\ \citenamefont
  {Hohenberg}}]{Werthamer1966}%
  \BibitemOpen
  \bibfield  {author} {\bibinfo {author} {\bibfnamefont {N.~R.}\ \bibnamefont
  {Werthamer}}, \bibinfo {author} {\bibfnamefont {E.}~\bibnamefont {Helfand}},
  \ and\ \bibinfo {author} {\bibfnamefont {P.~C.}\ \bibnamefont {Hohenberg}},\
  }\href {\doibase 10.1103/PhysRev.147.295} {\bibfield  {journal} {\bibinfo
  {journal} {Phys. Rev.}\ }\textbf {\bibinfo {volume} {147}},\ \bibinfo {pages}
  {295} (\bibinfo {year} {1966})}\BibitemShut {NoStop}%
\bibitem [{\citenamefont {Maki}(1964)}]{Maki_III}%
  \BibitemOpen
  \bibfield  {author} {\bibinfo {author} {\bibfnamefont {K.}~\bibnamefont
  {Maki}},\ }\href@noop {} {\bibfield  {journal} {\bibinfo  {journal}
  {Physics}\ }\textbf {\bibinfo {volume} {1}},\ \bibinfo {pages} {127}
  (\bibinfo {year} {1964})}\BibitemShut {NoStop}%
\bibitem [{\citenamefont {Hussey}\ \emph {et~al.}(2004)\citenamefont {Hussey},
  \citenamefont {Takenaka},\ and\ \citenamefont {Takagi}}]{Hussey2004}%
  \BibitemOpen
  \bibfield  {author} {\bibinfo {author} {\bibfnamefont {N.~E.}\ \bibnamefont
  {Hussey}}, \bibinfo {author} {\bibfnamefont {K.}~\bibnamefont {Takenaka}}, \
  and\ \bibinfo {author} {\bibfnamefont {H.}~\bibnamefont {Takagi}},\ }\href
  {\doibase 10.1080/14786430410001716944} {\bibfield  {journal} {\bibinfo
  {journal} {Philos. Mag.}\ }\textbf {\bibinfo {volume} {84}},\ \bibinfo
  {pages} {2847} (\bibinfo {year} {2004})}\BibitemShut {NoStop}%
\bibitem [{\citenamefont {Matsui}\ \emph {et~al.}(2007)\citenamefont {Matsui},
  \citenamefont {Takahashi}, \citenamefont {Sato}, \citenamefont {Terashima},
  \citenamefont {Ding}, \citenamefont {Uefuji},\ and\ \citenamefont
  {Yamada}}]{Matsui2007}%
  \BibitemOpen
  \bibfield  {author} {\bibinfo {author} {\bibfnamefont {H.}~\bibnamefont
  {Matsui}}, \bibinfo {author} {\bibfnamefont {T.}~\bibnamefont {Takahashi}},
  \bibinfo {author} {\bibfnamefont {T.}~\bibnamefont {Sato}}, \bibinfo {author}
  {\bibfnamefont {K.}~\bibnamefont {Terashima}}, \bibinfo {author}
  {\bibfnamefont {H.}~\bibnamefont {Ding}}, \bibinfo {author} {\bibfnamefont
  {T.}~\bibnamefont {Uefuji}}, \ and\ \bibinfo {author} {\bibfnamefont
  {K.}~\bibnamefont {Yamada}},\ }\href {\doibase 10.1103/PhysRevB.75.224514}
  {\bibfield  {journal} {\bibinfo  {journal} {Phys. Rev. B}\ }\textbf {\bibinfo
  {volume} {75}},\ \bibinfo {pages} {224514} (\bibinfo {year}
  {2007})}\BibitemShut {NoStop}%
\bibitem [{\citenamefont {Rieck}\ \emph {et~al.}(1991)\citenamefont {Rieck},
  \citenamefont {Scharnberg},\ and\ \citenamefont {Schopohl}}]{Rieck1991}%
  \BibitemOpen
  \bibfield  {author} {\bibinfo {author} {\bibfnamefont {C.~T.}\ \bibnamefont
  {Rieck}}, \bibinfo {author} {\bibfnamefont {K.}~\bibnamefont {Scharnberg}}, \
  and\ \bibinfo {author} {\bibfnamefont {N.}~\bibnamefont {Schopohl}},\ }\href
  {\doibase 10.1007/Bf00683526} {\bibfield  {journal} {\bibinfo  {journal} {J.
  Low Temp. Phys.}\ }\textbf {\bibinfo {volume} {84}},\ \bibinfo {pages} {381}
  (\bibinfo {year} {1991})}\BibitemShut {NoStop}%
\bibitem [{\citenamefont {Longuethiggins}\ \emph {et~al.}(1958)\citenamefont
  {Longuethiggins}, \citenamefont {\"Opik}, \citenamefont {Pryce},\ and\
  \citenamefont {Sack}}]{Longuethiggins1958}%
  \BibitemOpen
  \bibfield  {author} {\bibinfo {author} {\bibfnamefont {H.~C.}\ \bibnamefont
  {Longuethiggins}}, \bibinfo {author} {\bibfnamefont {U.}~\bibnamefont
  {\"Opik}}, \bibinfo {author} {\bibfnamefont {M.~H.~L.}\ \bibnamefont
  {Pryce}}, \ and\ \bibinfo {author} {\bibfnamefont {R.~A.}\ \bibnamefont
  {Sack}},\ }\href {\doibase 10.1098/rspa.1958.0022} {\bibfield  {journal}
  {\bibinfo  {journal} {Proc. Royal Soc. Lond.}\ }\textbf {\bibinfo {volume}
  {244}},\ \bibinfo {pages} {1} (\bibinfo {year} {1958})}\BibitemShut {NoStop}%
\bibitem [{\citenamefont {Khomskii}\ and\ \citenamefont
  {Mizokawa}(2005)}]{Khomskii2005}%
  \BibitemOpen
  \bibfield  {author} {\bibinfo {author} {\bibfnamefont {D.~I.}\ \bibnamefont
  {Khomskii}}\ and\ \bibinfo {author} {\bibfnamefont {T.}~\bibnamefont
  {Mizokawa}},\ }\href {\doibase 10.1103/PhysRevLett.94.156402} {\bibfield
  {journal} {\bibinfo  {journal} {Phys. Rev. Lett.}\ }\textbf {\bibinfo
  {volume} {94}},\ \bibinfo {pages} {156402} (\bibinfo {year}
  {2005})}\BibitemShut {NoStop}%
\bibitem [{\citenamefont {Watkins}\ \emph {et~al.}(2003)\citenamefont
  {Watkins}, \citenamefont {Mainwood},\ and\ \citenamefont
  {Davies}}]{Watkinsi2003}%
  \BibitemOpen
  \bibfield  {author} {\bibinfo {author} {\bibfnamefont {M.}~\bibnamefont
  {Watkins}}, \bibinfo {author} {\bibfnamefont {A.}~\bibnamefont {Mainwood}}, \
  and\ \bibinfo {author} {\bibfnamefont {G.}~\bibnamefont {Davies}},\ }\href
  {\doibase 10.1016/S0925-9635(02)00235-2} {\bibfield  {journal} {\bibinfo
  {journal} {Diam. Relat. Mater.}\ }\textbf {\bibinfo {volume} {12}},\ \bibinfo
  {pages} {503} (\bibinfo {year} {2003})}\BibitemShut {NoStop}%
\bibitem [{\citenamefont {Haverkort}\ \emph {et~al.}(2005)\citenamefont
  {Haverkort}, \citenamefont {Hu}, \citenamefont {Tanaka}, \citenamefont
  {Ghiringhelli}, \citenamefont {Roth}, \citenamefont {Cwik}, \citenamefont
  {Lorenz}, \citenamefont {Sch\"u$\beta$ler-Langeheine}, \citenamefont
  {Streltsov}, \citenamefont {Mylnikova}, \citenamefont {Anisimov},
  \citenamefont {de~Nadai}, \citenamefont {Brookes}, \citenamefont {Hsieh},
  \citenamefont {Lin}, \citenamefont {Chen}, \citenamefont {Mizokawa},
  \citenamefont {Taguchi}, \citenamefont {Tokura}, \citenamefont {Khomskii},\
  and\ \citenamefont {Tjeng}}]{Haverkort2005}%
  \BibitemOpen
  \bibfield  {author} {\bibinfo {author} {\bibfnamefont {M.~W.}\ \bibnamefont
  {Haverkort}}, \bibinfo {author} {\bibfnamefont {Z.}~\bibnamefont {Hu}},
  \bibinfo {author} {\bibfnamefont {A.}~\bibnamefont {Tanaka}}, \bibinfo
  {author} {\bibfnamefont {G.}~\bibnamefont {Ghiringhelli}}, \bibinfo {author}
  {\bibfnamefont {H.}~\bibnamefont {Roth}}, \bibinfo {author} {\bibfnamefont
  {M.}~\bibnamefont {Cwik}}, \bibinfo {author} {\bibfnamefont {T.}~\bibnamefont
  {Lorenz}}, \bibinfo {author} {\bibfnamefont {C.}~\bibnamefont
  {Sch\"u$\beta$ler-Langeheine}}, \bibinfo {author} {\bibfnamefont {S.~V.}\
  \bibnamefont {Streltsov}}, \bibinfo {author} {\bibfnamefont {A.~S.}\
  \bibnamefont {Mylnikova}}, \bibinfo {author} {\bibfnamefont {V.~I.}\
  \bibnamefont {Anisimov}}, \bibinfo {author} {\bibfnamefont {C.}~\bibnamefont
  {de~Nadai}}, \bibinfo {author} {\bibfnamefont {N.~B.}\ \bibnamefont
  {Brookes}}, \bibinfo {author} {\bibfnamefont {H.~H.}\ \bibnamefont {Hsieh}},
  \bibinfo {author} {\bibfnamefont {H.~J.}\ \bibnamefont {Lin}}, \bibinfo
  {author} {\bibfnamefont {C.~T.}\ \bibnamefont {Chen}}, \bibinfo {author}
  {\bibfnamefont {T.}~\bibnamefont {Mizokawa}}, \bibinfo {author}
  {\bibfnamefont {Y.}~\bibnamefont {Taguchi}}, \bibinfo {author} {\bibfnamefont
  {Y.}~\bibnamefont {Tokura}}, \bibinfo {author} {\bibfnamefont {D.~I.}\
  \bibnamefont {Khomskii}}, \ and\ \bibinfo {author} {\bibfnamefont {L.~H.}\
  \bibnamefont {Tjeng}},\ }\href {\doibase 10.1103/Physrevlett.94.056401}
  {\bibfield  {journal} {\bibinfo  {journal} {Phys. Rev. Lett.}\ }\textbf
  {\bibinfo {volume} {94}},\ \bibinfo {pages} {056401} (\bibinfo {year}
  {2005})}\BibitemShut {NoStop}%
\bibitem [{\citenamefont {Bersuker}(2006)}]{Bersuker2006}%
  \BibitemOpen
  \bibfield  {author} {\bibinfo {author} {\bibfnamefont {I.~B.}\ \bibnamefont
  {Bersuker}},\ }\href@noop {} {\emph {\bibinfo {title} {The Jahn-Teller
  effect}}}\ (\bibinfo  {publisher} {Cambridge University Press},\ \bibinfo
  {address} {Cambridge},\ \bibinfo {year} {2006})\BibitemShut {NoStop}%
\bibitem [{\citenamefont {Mizokawa}\ and\ \citenamefont
  {Fujimori}(1996)}]{Mizokawa1996}%
  \BibitemOpen
  \bibfield  {author} {\bibinfo {author} {\bibfnamefont {T.}~\bibnamefont
  {Mizokawa}}\ and\ \bibinfo {author} {\bibfnamefont {A.}~\bibnamefont
  {Fujimori}},\ }\href {\doibase 10.1103/PhysRevB.54.5368} {\bibfield
  {journal} {\bibinfo  {journal} {Phys. Rev. B}\ }\textbf {\bibinfo {volume}
  {54}},\ \bibinfo {pages} {5368} (\bibinfo {year} {1996})}\BibitemShut
  {NoStop}%
\bibitem [{\citenamefont {Meijer}\ \emph {et~al.}(1999)\citenamefont {Meijer},
  \citenamefont {Henggeler}, \citenamefont {Brown}, \citenamefont {Becker},
  \citenamefont {Bednorz}, \citenamefont {Rossel},\ and\ \citenamefont
  {Wachter}}]{Meijer1999}%
  \BibitemOpen
  \bibfield  {author} {\bibinfo {author} {\bibfnamefont {G.~I.}\ \bibnamefont
  {Meijer}}, \bibinfo {author} {\bibfnamefont {W.}~\bibnamefont {Henggeler}},
  \bibinfo {author} {\bibfnamefont {J.}~\bibnamefont {Brown}}, \bibinfo
  {author} {\bibfnamefont {O.~S.}\ \bibnamefont {Becker}}, \bibinfo {author}
  {\bibfnamefont {J.~G.}\ \bibnamefont {Bednorz}}, \bibinfo {author}
  {\bibfnamefont {C.}~\bibnamefont {Rossel}}, \ and\ \bibinfo {author}
  {\bibfnamefont {P.}~\bibnamefont {Wachter}},\ }\href {\doibase
  10.1103/PhysRevB.59.11832} {\bibfield  {journal} {\bibinfo  {journal} {Phys.
  Rev. B}\ }\textbf {\bibinfo {volume} {59}},\ \bibinfo {pages} {11832}
  (\bibinfo {year} {1999})}\BibitemShut {NoStop}%
\bibitem [{\citenamefont {Neuringer}\ and\ \citenamefont
  {Shapira}(1966)}]{Neuringer1966}%
  \BibitemOpen
  \bibfield  {author} {\bibinfo {author} {\bibfnamefont {L.~J.}\ \bibnamefont
  {Neuringer}}\ and\ \bibinfo {author} {\bibfnamefont {Y.}~\bibnamefont
  {Shapira}},\ }\href {\doibase 10.1103/PhysRevLett.17.81} {\bibfield
  {journal} {\bibinfo  {journal} {Phys. Rev. Lett.}\ }\textbf {\bibinfo
  {volume} {17}},\ \bibinfo {pages} {81} (\bibinfo {year} {1966})}\BibitemShut
  {NoStop}%
\bibitem [{\citenamefont {Boiko}\ and\ \citenamefont
  {Rashba}(1960)}]{Rashba1960}%
  \BibitemOpen
  \bibfield  {author} {\bibinfo {author} {\bibfnamefont {I.~I.}\ \bibnamefont
  {Boiko}}\ and\ \bibinfo {author} {\bibfnamefont {E.~I.}\ \bibnamefont
  {Rashba}},\ }\href@noop {} {\bibfield  {journal} {\bibinfo  {journal} {Fizika
  Tverdogo Tela}\ }\textbf {\bibinfo {volume} {2}},\ \bibinfo {pages} {1874}
  (\bibinfo {year} {1960})}\BibitemShut {NoStop}%
\bibitem [{\citenamefont {Gor$^{\prime}$kov}\ and\ \citenamefont
  {Rusinov}(1964)}]{Gorkov1964}%
  \BibitemOpen
  \bibfield  {author} {\bibinfo {author} {\bibfnamefont {L.~P.}\ \bibnamefont
  {Gor$^{\prime}$kov}}\ and\ \bibinfo {author} {\bibfnamefont {A.~I.}\
  \bibnamefont {Rusinov}},\ }\href@noop {} {\bibfield  {journal} {\bibinfo
  {journal} {JETP}\ }\textbf {\bibinfo {volume} {19}},\ \bibinfo {pages} {922}
  (\bibinfo {year} {1964})}\BibitemShut {NoStop}%
\bibitem [{\citenamefont {Zhou}\ \emph {et~al.}(2016)\citenamefont {Zhou},
  \citenamefont {Yuan}, \citenamefont {Jiang},\ and\ \citenamefont
  {Law}}]{Zhou2016}%
  \BibitemOpen
  \bibfield  {author} {\bibinfo {author} {\bibfnamefont {B.~T.}\ \bibnamefont
  {Zhou}}, \bibinfo {author} {\bibfnamefont {N.~F.~Q.}\ \bibnamefont {Yuan}},
  \bibinfo {author} {\bibfnamefont {H.~L.}\ \bibnamefont {Jiang}}, \ and\
  \bibinfo {author} {\bibfnamefont {K.~T.}\ \bibnamefont {Law}},\ }\href
  {\doibase 10.1103/PhysRevB.93.180501} {\bibfield  {journal} {\bibinfo
  {journal} {Phys. Rev. B}\ }\textbf {\bibinfo {volume} {93}},\ \bibinfo
  {pages} {180501} (\bibinfo {year} {2016})}\BibitemShut {NoStop}%
\bibitem [{\citenamefont {Kusmartsev}\ and\ \citenamefont
  {Tsvelik}(1985)}]{Kusmartsev1985}%
  \BibitemOpen
  \bibfield  {author} {\bibinfo {author} {\bibfnamefont {F.~V.}\ \bibnamefont
  {Kusmartsev}}\ and\ \bibinfo {author} {\bibfnamefont {A.~M.}\ \bibnamefont
  {Tsvelik}},\ }\href@noop {} {\bibfield  {journal} {\bibinfo  {journal}
  {JETP}\ }\textbf {\bibinfo {volume} {42}},\ \bibinfo {pages} {257} (\bibinfo
  {year} {1985})}\BibitemShut {NoStop}%
\bibitem [{\citenamefont {Pesin}\ and\ \citenamefont
  {Balents}(2010)}]{Dmytro2010}%
  \BibitemOpen
  \bibfield  {author} {\bibinfo {author} {\bibfnamefont {D.}~\bibnamefont
  {Pesin}}\ and\ \bibinfo {author} {\bibfnamefont {L.}~\bibnamefont
  {Balents}},\ }\href {\doibase 10.1038/Nphys1606} {\bibfield  {journal}
  {\bibinfo  {journal} {Nat. Phys.}\ }\textbf {\bibinfo {volume} {6}},\
  \bibinfo {pages} {376} (\bibinfo {year} {2010})}\BibitemShut {NoStop}%
\bibitem [{\citenamefont {Galitski}\ and\ \citenamefont
  {Spielman}(2013)}]{Galitski2013}%
  \BibitemOpen
  \bibfield  {author} {\bibinfo {author} {\bibfnamefont {V.}~\bibnamefont
  {Galitski}}\ and\ \bibinfo {author} {\bibfnamefont {I.~B.}\ \bibnamefont
  {Spielman}},\ }\href {\doibase 10.1038/nature11841} {\bibfield  {journal}
  {\bibinfo  {journal} {Nature}\ }\textbf {\bibinfo {volume} {494}},\ \bibinfo
  {pages} {49} (\bibinfo {year} {2013})}\BibitemShut {NoStop}%
\bibitem [{\citenamefont {Rykov}\ \emph {et~al.}(1999)\citenamefont {Rykov},
  \citenamefont {Tajima}, \citenamefont {Kusmartsev}, \citenamefont {Forgan},\
  and\ \citenamefont {Simon}}]{Rykov1999}%
  \BibitemOpen
  \bibfield  {author} {\bibinfo {author} {\bibfnamefont {A.~I.}\ \bibnamefont
  {Rykov}}, \bibinfo {author} {\bibfnamefont {S.}~\bibnamefont {Tajima}},
  \bibinfo {author} {\bibfnamefont {F.~V.}\ \bibnamefont {Kusmartsev}},
  \bibinfo {author} {\bibfnamefont {E.~M.}\ \bibnamefont {Forgan}}, \ and\
  \bibinfo {author} {\bibfnamefont {C.}~\bibnamefont {Simon}},\ }\href
  {\doibase 10.1103/PhysRevB.60.7601} {\bibfield  {journal} {\bibinfo
  {journal} {Phys. Rev. B}\ }\textbf {\bibinfo {volume} {60}},\ \bibinfo
  {pages} {7601} (\bibinfo {year} {1999})}\BibitemShut {NoStop}%
\end{thebibliography}%

\end{document}